\documentclass{article}
\usepackage{spconf,amsmath, amssymb, graphicx}
\DeclareGraphicsExtensions{.pdf,.png,.jpg}
\usepackage{makecell,multirow}
\usepackage{adjustbox}      
\usepackage{subfig}
\usepackage{hhline}
\usepackage{amsmath}
\usepackage[font=small,labelfont=bf]{caption}
\usepackage{amsmath}
\usepackage{algorithm}
\usepackage[noend]{algpseudocode}

\usepackage{kotex}
\usepackage{xcolor}

\makeatletter
\def\BState{\State\hskip-\ALG@thistlm}
\makeatother

\title{Listen to Dance: Music-driven choreography generation \\ using Autoregressive Encoder-Decoder Network}
\name{Juheon Lee, Seohyun Kim, Kyogu Lee}
\address{ Music \& Audio Research Group\\
Seoul National University, Seoul, Korea}
\begin{document}
%
\maketitle
\begin{abstract}
Automatic choreography generation is a challenging task because it often requires an understanding of two abstract concepts - music and dance - which are realized in the two different modalities, namely audio and video, respectively. In this paper, we propose a music-driven choreography generation system using an auto-regressive encoder-decoder network. To this end, we first collect a set of multimedia clips that include both music and corresponding dance motion. We then extract the joint coordinates of the dancer from video and the mel-spectrogram of music from audio, and train our network using music-choreography pairs as input. Finally, a novel dance motion is generated at the inference time when only music is given as an input. We performed a user study for a qualitative evaluation of the proposed method, and the results show that the proposed model is able to generate musically meaningful and natural dance movements given an unheard song.




\end{abstract}
\begin{keywords}
Choreography, dance motion generation, autoregressive encoder-decoder network
\end{keywords}
\section{Introduction}
\label{sec:intro}

Choreography is a kind of art that designs a series of movements. In particular, in performing art, choreography extends to the use of human bodies to express movements, and these are often performed with music. The choreography suitable for music has significance in that it is not only an artwork itself, but also maximizes the expression of music. \cite{krumhansl1997can, gentry2004modeling} For this reason, choreography has become an essential element in many pop music works in recent years. Therefore, the process of creating choreography for music is also considered to be important, and research on a system capable of automatically generating choreography is actively conducted. However, automatic choreography generation is a challenging task because both music and dance are abstract art concepts, and the clear relationship between the two concepts is also not defined by established rules.

   

Recent advances in machine learning and deep learning techniques have led to a variety of attempts to study the relationship between dance and music.  Lee et al. proposed a choreography generation algorithm that retrieves the motions corresponding to the most similar pieces of music in the predefined motion-music-paired database for given new music segment. \cite{lee2013music}. This method selects dance motion from a predefined database, so choreography retrieved with high correlation with music is guaranteed. However, it has limitations in that it can not create novel dance movements that are not included in the database. Ofil et al. proposed a HMM-based model that categorizes the genre of music based on the Mel-Frequency Cepstrum Coefficients (MFCC) feature and generates matching choreography based on the results \cite{ofli2008audio}. But since the choreography is determined by the categorical value obtained through the genre classifier, there is a limit to generate a novel choreography. Omid et al. proposed a music-driven choreography model named Groovenet \cite{alemi2017groovenet}. They used pairs of music and three-dimensional motion data to train the Factored Conditional Restricted Boltzmann
Machines (FCRBM) \cite{taylor2009factored}. They attempted to directly train the relationship between music and dance by using the mel spectrogram in the training process. However, they reported that their model created awkward dance moves for unheard song, so they conclude that the model was overfitted and the dance moves according to music were not generalized enough.

\begin{figure*}[ht]
\centering
\includegraphics[width=0.9\linewidth]{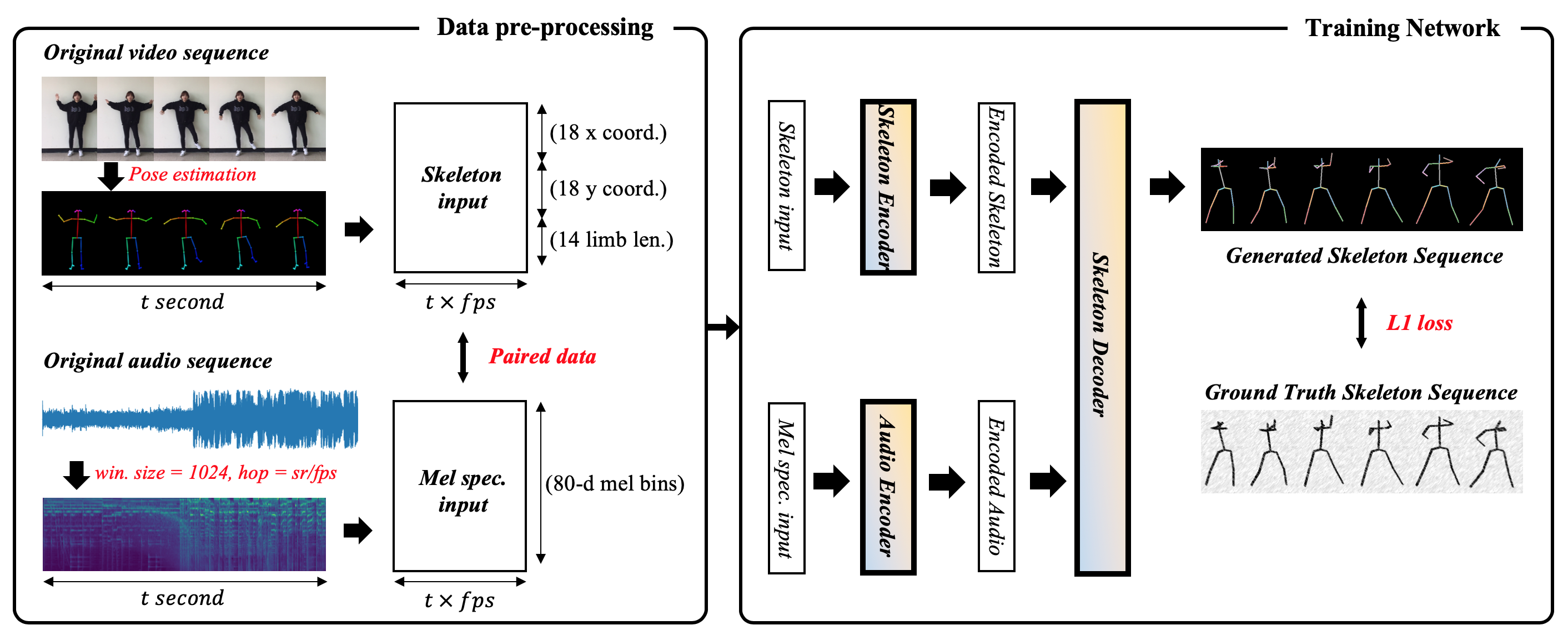}
\caption{
A schematic diagram of the proposed music-driven choreography generation system.} 
\label{fig:network}
\end{figure*}

Lee et al.'s and Ofil et al.'s studies have a limitation in that they can not create novel choreography because the former synthesizes motion by reusing the choreographic samples in a predefined database, and the latter creates choreography only for music input categorized by its genre. Omid et al. did succeed to create novel dance motion, but failed to yield good results mainly due to insufficient training data of merely 23 minutes. In this study, we propose a neural network-based model that can generate novel and natural choreography trained on large amount of data that is easily obtained from the online video sharing community. An overview of the proposed system is illustrated in Figure \ref{fig:network}.

The rest of the paper is organized as follows. In Section \ref{sec:cross}, we explain in detail our proposed method for choreography generation based on the encoder-decoder network. We describe the datasets for experiments and the training process in Section \ref{sec:exp}. The evaluation scheme and the results are presented in Section \ref{sec:eval}, followed by conclusions and directions for future work in Section \ref{sec:con}.



\section{Listen to Dance: proposed approach}
\label{sec:cross}

In order to learn the relationship between the time-series data of two different modalities, i.e., music and dance, we need a model that performs multi-modal sequence-to-sequence transformations. We modified the Dilated Convolution Text-To-Speech model \cite{tachibana2017efficiently} that performed well in the text-to-speech domain and used it as a our choreography generation network. Our model consists of two encoders and one decoder, and the detailed structure is explained below.

\subsection{Causal Dilated Highway Conv. Block}

The encoders and decoder used in the proposed model contain causal dilated highway convolutional blocks (CDHC). \textit{Causal} means that only the input data from time $0$ to $t-1$ can be referred to when calculating the output at time $t$. We used a causal convolution layer because our network must be an auto-regressive model to generate the next frame that is not yet known from the preceding frames. In addition, we used the \textit{dilated convolution} used in the Wavenet \cite{van2016wavenet} to ensure that the model has a wider receptive field. Finally, to enable efficient training even in deep model structures, we used a \textit{highway network} architecture \cite{srivastava2015highway} where gated function could be trained. That is, the output of the CDHC block is calculated as:
\begin{equation}
    \textbf{output} = tanh(\text{H1}) \cdot relu(\text{H2}) + (1-tanh(\text{H1})) \cdot \textbf{input}
\end{equation}

[H1, H2] is the tensor calculated through the causal dilated convolution layer of the input tensor. The output channel of this convolution layer is twice the input channel, and the kernel size is 3.

\subsection{Encoder \& Decoder}

Both the skeleton encoders and the audio encoders all consist of three convolution layers and 10 CDHC blocks. The first convolution layer of each encoder increases the input channel to 256 dimensions, and the other two layers perform 1x1 convolution. Thereafter, the output values from last convolutional layer are connected in sequence to 10 CDHC blocks with a dilation factor of (1,3,9,27,1,3,9,27,3,3), and the corresponding operations result in audio and skeleton data are encoded to have a sufficiently wide receptive field to reflect sufficient past information. 

A decoder is a network that generates skeleton data for the next frame from an encoded skeleton and an encoded audio. First, the encoded skeleton input to the decoder is combined with the encoded audio in the following:

\begin{equation}
    H1 = conv(\textbf{$E_{skeleton}$}) + E_{audio}[:128]
\end{equation}
\begin{equation}
    H2 = conv(\textbf{$E_{skeleton}$}) + E_{audio}[128:]
\end{equation}
\begin{equation}
    \textbf{comb} = \sigma(H1) \cdot tanh(H2)
\end{equation}

Where $E_{skeleton}$ and $E_{audio}$ refer to the encoded skeleton and encoded audio, respectively, and $conv$ means the convolution layer with an output channel of 128 and a kernel size of 1. The combined \textbf{comb} tensor then goes through six CDHC blocks with a dilation factor of (1,3,9,27,3,3) and then through three 128-channel convolutional layers with a \textbf{tanh} activation function. Finally, after passing through a convolution layer with the same output channel as the dimension of the target, the final decoder output is obtained via \textbf{sigmoid} activation.

\subsection{Proposed network}
This network receives motion and music data from time $0$ to $t-1$ as input. Both data are encoded via encoders and combined at the beginning of the decoder. The final output of the decoder is compared with the ground truth motion data at time $1$ to $t$ and we used it as a $L1$ loss. Since all convolution operations included in the network are with kernel size 1 or causal operations, the $k$th value of output refers to only the $0$ to $k-1$ time step of the input during the operation. Therefore, the model satisfies the auto-regressive condition.

\section{Experiment}
\label{sec:exp}

\subsection{Data}

We have collected 100 YouTube choreography videos and corresponding audios. The genre was selected mainly for K-pop dance, and the total length of collected data was 6.26 hours.

\subsubsection{Skeleton data}
We extracted the x, y coordinates of 15 human body joints from each frame using the Openpose algorithm \cite{cao2017realtime} from the collected video as shown in Fig. 2. Next, we min-max normalize the extracted coordinate values for each video, and use the linear-interpolation for the unrecognized coordinate values.

Since we can not measure the exact 3d angle between the human body limbs using the 2d joint coordinate, we used the absolute coordinates values of each point as the training target. However, in this case, the length of each limb in the projected skeleton can vary, and awkward motion can be generated if the model learns it incorrectly. So we additionally calculated the lengths of the 14 main limbs together and added a loss to compare with the limb length of the skeleton that the model generated. Therefore, the x, y coordinates of the total 15 joints, and the total of 14 main limb length are used as skeleton data.


\subsubsection{Music data}
We separated the audio contained in the collected video and used it as music data. The mel-spectrogram was extracted from the audio waveform with the window size of 1024 samples, and 80 mel-frequency bins. Because we need time-aligned audio-video pairs for training, we adjusted the hop size when extracting the mel-spectrogram so that audio data has the same frame rate as that of video.

\subsection{Training}
We have trained proposed network that creates the next skeleton coordinate for a given previous skeleton sequence and music sequence. To do this, we first input skeleton data and music data from 0 to t-1 frames. Then, the output of the network is compared with the ground truth choreographic data corresponding to 1 to t frame by use L1 loss as a cost function. In addition, we calculated the length of each limb from the skeleton data of the generated frame, and compared with the actual ground truth length through the L1 loss. We used the adam optimizer \cite{kingma2014adam} for training and set the learning rate to 0.0002.


\begin{figure}[t]
\includegraphics[width=0.95\linewidth]{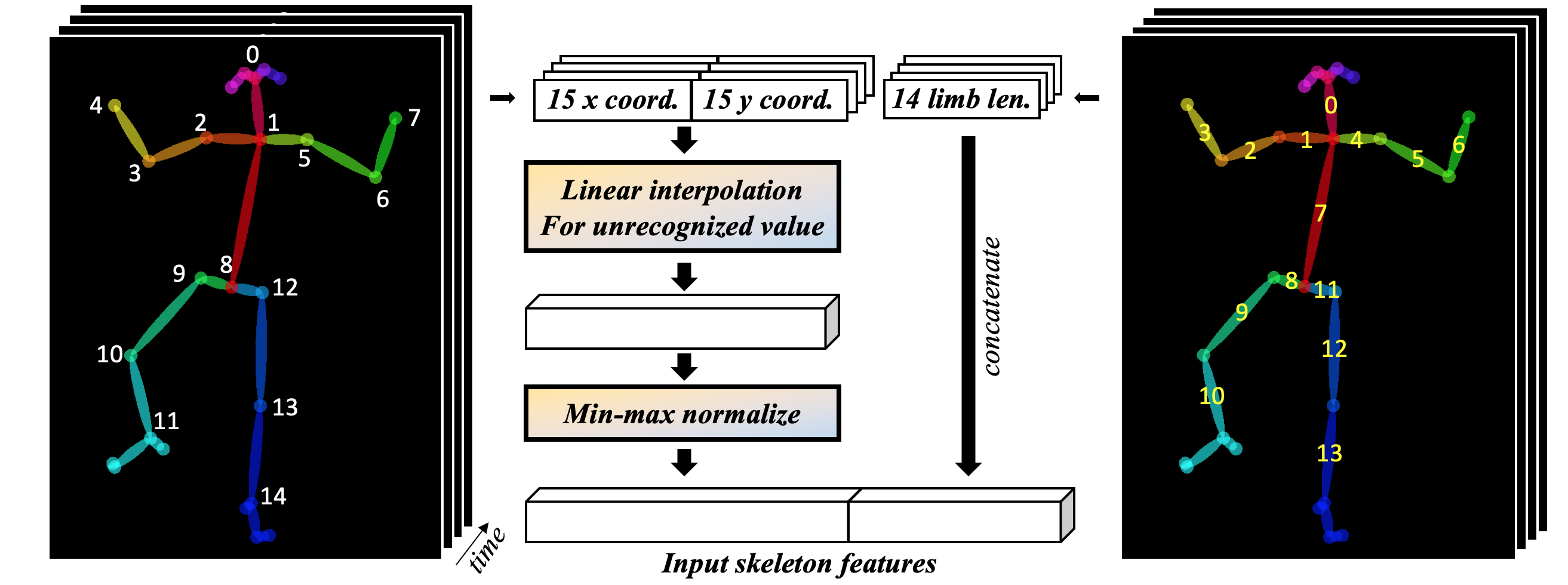}
\caption{The process of extracting skeleton data from video frames.} 
\label{fig:dataset}
\end{figure}

\subsection{Inference}
The choreography inference process is performed in an auto-regressive manner different from training. That is, the initial position of each joint is given as an input skeleton frame, and at the same time, the first frame of mel spectrogram is input to the trained model. When inference is performed once, estimated skeleton at $t=1$ is output. Then we concatenate skeleton at $t=0$ and $t=1$, then input them back into the model with mel spectrogram at $t=0$ and $t=1$ . After than, we get estimated skeleton at $t=1$ and $t=2$. Therefore, we can generate the choreography by repeating the above process for the length of music input, and used it to evaluate the generated choreography.

\section{Evaluation \& Results}
\label{sec:eval}

\subsection{User study}
We conducted a user study to evaluate whether the generated choreography was natural and whether it was produced in accordance with the music. First, we generated 20 videos for each of the three groups: \textit{Real}, \textit{Generated}, and \textit{Random}. Group \textit{Real} consists of music $A_i$ and actual choreography for music $A_i$. Group \textit{Generated} consists of music $B_i$ and novel choreography generated by our model given music $B_i$. Finally, the group \textit{Random} consists of music $C_i$ and novel choreography generated by our model but with randomly selected music rather than $C_i$. Music $A_i$, $B_i$, and $C_i$ were randomly selected among the songs included in the validation dataset that was not used in training, and the length of each audio/video was 16 seconds. \footnote{
The generated result can be found at: listentodance.strikingly.com.}

After mixing the three groups of videos in a random order, we asked the participants whether each video's choreography is natural (Question 1) and whether it fits well with music (Question 2), and to give a score in a Likert scale \cite{allen2007likert}. After collecting the responses, we performed isoquantity and normality tests using data averaging 20 responses from each group, to see if there was a difference in the mean of the responses of the groups. After evaluating significance through repeated-measure ANOVA test , further post-hoc analysis was performed to calculate the p-value, and the difference between the groups was examined \cite{girden1992anova}.

A total of 33 participants answered the questionnaire and the results are shown in the Fig.3. The results of the statistical tests confirmed that the mean scores between the three groups were significantly different for both questions. Average user score for both questions were highest in \textit{Real} group and lowest in \textit{Random} group. It is clear that the \textit{Real} group score is the highest, because it is made up of the choreography created by the human. The average score of the \textit{Generated} group surpassed the \textit{Random} group in both questions. If the proposed model generates choreography that is not associated with music, participants will have a similar response, regardless of what music is played with the generated choreography. However, from the fact that the video received a significantly higher score when played with the music used in choreography generation, we judged that the proposed model produced choreography that listen and reflects the music.

\begin{figure}
    \centering
    \includegraphics[width=0.45\textwidth]{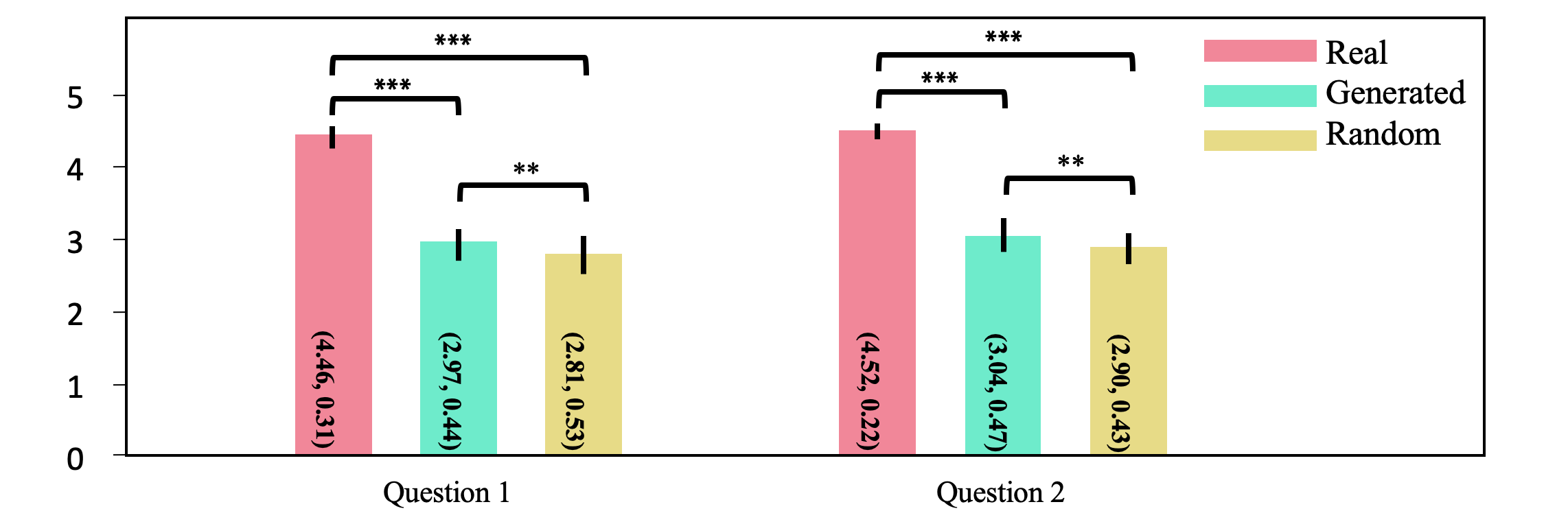}
    
    \caption{Average Likert-scale user scores on two questions (Q1: Is choreography natural? / Q2: Does choreography fit well with music?). The numbers on the bar graph are the mean score and standard deviation of the responses, respectively. The p-values for pairwise comparisons between the groups are also shown at the top. ***: $p < 0.001$; **: $p < 0.01$.}
    \label{fig:my_label}
\end{figure}


\subsection{Autocorrelation Analysis}

We also performed an autocorrelation analysis to further investigate the differences between the generated choreography and the actual choreography. Autocorrelation is a correlation between a given sequence with itself, reflecting the periodic properties of the sequence. We can identify the periodic component of a given sequence through the location of the peaks observed in the autocorrelation results. Using this, we analyzed the motion by calculating the autocorrelation on the x, y coordinates of the choreography movement and compared it with the tempo of corresponding music. Our hypothesis was that if the model can produce dance by listening to the music, the autocorrelation peak position of the motion will appear at the same point as the beat of the music.

Fig. 4 shows the autocorrelation results of two choreography samples along with the tempo of corresponding music. In actual choreography, a clear peak is observed in y-direction movement, but not in x-direction movement. This tendency is also observed in the generated choreography. From this we can determine that the proposed network has learned the periodic tendency of the real choreography used in training. Also, In actual choreography, the first or second peak of the y-direction auto-correlation appears at the same position as the music beat. This means that music and choreography have similar periodic properties. This tendency can be confirmed also in the case of the generated sample. From this, it is judged that the proposed model has generated the choreography that listen the music and reflects its periodic nature.

\begin{figure}
    \centering
    \includegraphics[width=0.5\textwidth]{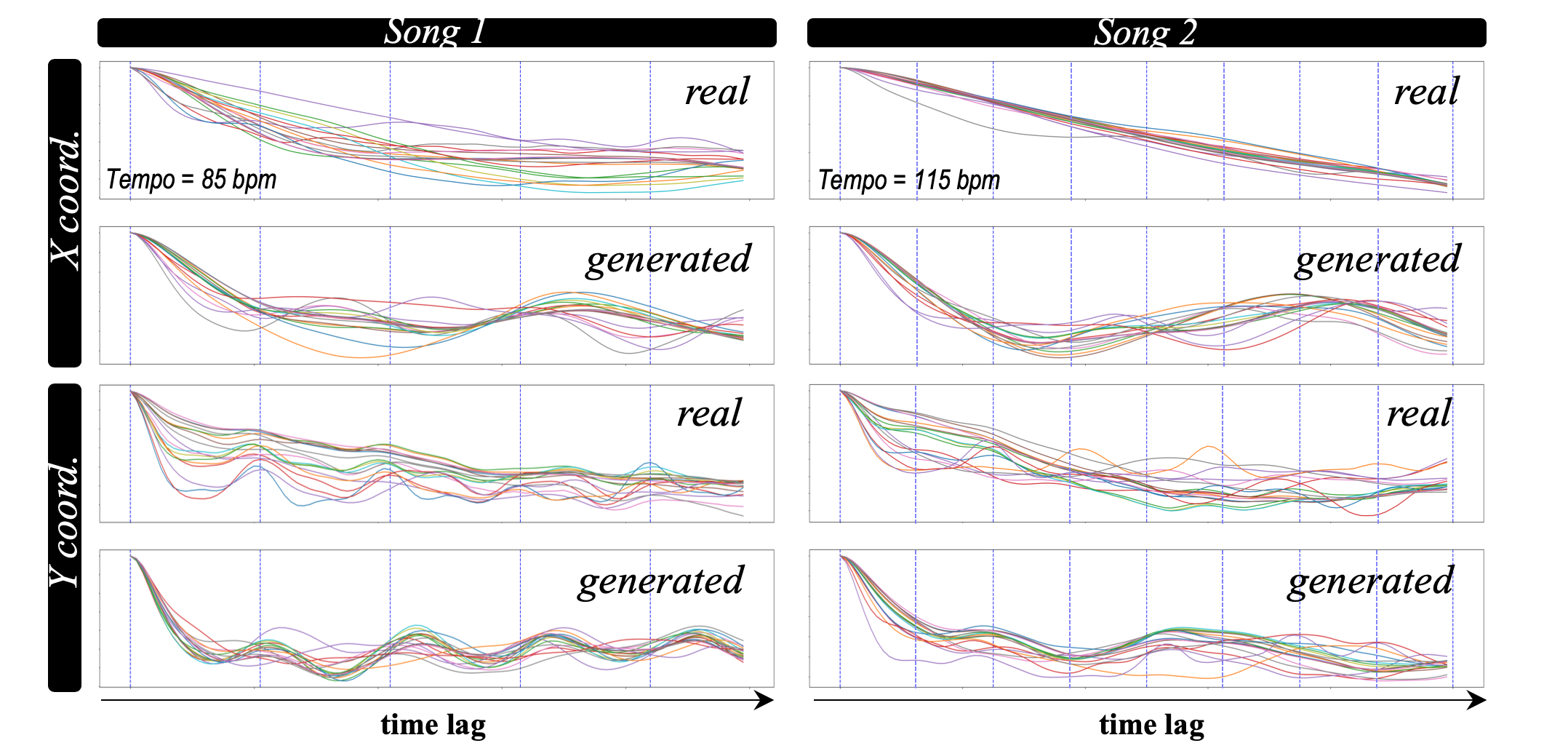}
    \caption{Autocorrelation of the x,y coordinates from real and generated choreography for two songs. The x-axis of each graph represents the time lag, and the blue vertical lines represent the beat positions of each song.}
    \label{fig:my_label}
\end{figure}

\section{conclusion}
\label{sec:con}

In this study, we proposed an auto-regressive encoder-decoder network that generates matching choreography for a given music input. We used audio-video pairs data obtained from YouTube for training. As a result, it was found that motions matching with the music were generated through comparison of user study and autocorrelation analysis. This study has a significance in that it shows a significant performance in the area of learning-based choreography generation, in which sufficient performance has not been secured yet. Also, it is meaningful not only to learn the movement of dance but also to use the relationship with music together for generation.

This research has limitations that generated choreography reflects only the periodicity among various properties of music. Ultimately, it is necessary to create appropriate choreography according to various genres, moods, and contexts of music as well as periodicity. In order to do this, we plan to establish data sets that satisfy various conditions and carry out further research. In addition, we use 2-d skeleton position for training, and it is difficult to use this type of data in case of needing actual implementation such as a robot. Therefore, the extension of the model to 3-d choreography generation using the improved 3-d pose estimation algorithm is also a future research topic.

\section{Acknowledgements}
This work was supported by LG electronics.\\

\bibliographystyle{IEEEbib.bst}
\bibliography{bibliography.bib}

\end{document}